\documentclass[twocolumn,showpacs,preprintnumbers,amsmath,amssymb]{revtex4}

\usepackage{graphicx}
\usepackage{dcolumn}
\usepackage{bm}

\begin{document}

\title{Quantification of the Heterogeneity of Particle Packings}

\author{Iwan Schenker}
 \email{iwan.schenker@alumni.ethz.ch}
\author{Frank T. Filser}%
\author{Ludwig J. Gauckler}%
\affiliation{ Nonmetallic Materials, Department of Materials, ETH
Zurich, Zurich CH-8093, Switzerland
\homepage{http://www.ceramics.ethz.ch}
}%

\author{Tomaso Aste}
\affiliation{Department of Applied Mathematics, RSPhysSE, The
Australian National University, 0200 Australia\\and\\School of
Physical Sciences, University of Kent, Canterbury, Kent, CT2 7NH,
United Kingdom}%

\author{Hans J. Herrmann}
\affiliation{ Computational Physics for Engineering Materials,
Institute for Building Materials, ETH Zurich, Zurich CH-8093,
Switzerland
}%


\begin{abstract}

The microstructure of coagulated colloidal particles, for which
the inter-particle potential is described by the DLVO theory, is
strongly influenced by the particles' surface potential. Depending
on its value, the resulting microstructures are either more
"homogeneous" or more "heterogeneous", at equal volume fractions.
An adequate quantification of a structure's degree of
heterogeneity (DOH) however does not yet exist. In this work,
methods to quantify and thus classify the DOH of microstructures
are investigated and compared. Three methods are evaluated using
particle packings generated by Brownian dynamics simulations: (1)
the pore size distribution, (2) the density fluctuation method and
(3) the Voronoi volume distribution. Each method provides a scalar
measure, either via a parameter in a fit function or an integral,
which correlates with the heterogeneity of the microstructure and
which thus allows for the first time to quantitatively capture the
DOH of a granular material. An analysis of the differences in the
density fluctuations between two structures additionally allows
for a detailed determination of the length scale on which
differences in heterogeneity are most pronounced.

\end{abstract}


\maketitle

\section{\label{sec:Intro}Introduction}

Colloidal particle packings are suitable model systems for the
study of the structural properties of granular materials below the
random loose packing limit. For such systems the gravitational
force is negligible in comparison to the van der Waals force,
electrostatic repulsion or Brownian
motion~\cite{Visser_1989,Israelachvili_1991}. In the present
study, we particularly focus on systems, for which the local
arrangement of the particles is the only variable, as opposed to
variations in the volume fraction or the particle size
distribution for example. Commonly, these microstructures are
referred to as "more homogeneous" or "more heterogeneous", which
either designates a structure presenting a rather uniform
distribution of the particle positions or one having locally
denser regions and therefore larger voids. These qualitative terms
may be intuitive, however, they do not allow for a sound
scientific quantification of the structure's degree of
heterogeneity (DOH), which does not yet exist. In this paper,
three methods providing the means for such a quantification are
presented, analyzed and compared. These methods permit for the
first time to explicitly capture the DOH of a particle packing in
form of a quantitative, scalar measure.

Experimentally, the reproducible generation of colloidal particle
packings possessing a specific DOH is achieved by the use of an
in-situ enzyme-catalyzed destabilization method (Direct
Coagulation Casting (DCC),~\cite{Gauckler_1999,Tervoort_2004}).
For volume fractions between 0.2 and 0.6, DCC allows for the
coagulation of electrostatically stabilized colloidal suspensions
to stiff particle structures by either shifting the pH of the
suspension to the particles' isoelectric point or by increasing
the ionic strength of the suspension without disturbing the
particle system. Shifting the pH leads to "more homogeneous"
microstructure through diffusion limited aggregation while
increasing the ionic strength results in "more heterogeneous"
microstructures via reaction rate limited aggregation. The
differences in heterogeneity have been observed using various
techniques such as diffusing wave spectroscopy~\cite{Wyss_2001},
static light transmission~\cite{Wyss_2001} or cryogenic scanning
electron microscopy~\cite{Wyss_2002}.

\begin{figure}
\includegraphics[width=\linewidth]{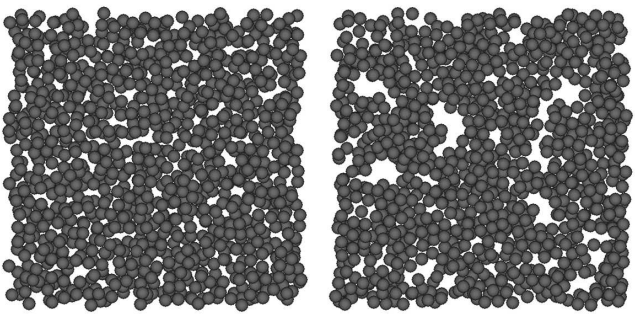}
\caption{\label{fig:fig_1} Slices through a "homogeneous" (left)
and a "heterogeneous" (right) particle structure with the same
volume fraction of 0.4 resulting from Brownian dynamics
simulations~\cite{Huetter_2000} (slice thickness: three particle
diameters; particle diameter: 0.5 $\mu$m).}
\end{figure}

Computationally, microstructures with different DOH were
successfully reproduced using Brownian dynamics simulations
(BD)~\cite{Wyss_2002,Huetter_2000,Huetter_1999}. Slices of three
particle layer thickness through a "homogeneous" and a
"heterogeneous" BD-microstructure of identical volume fraction
nicely demonstrate the differences between the microstructures
(Fig.~\ref{fig:fig_1}). The microstructure on the left presents a
rather uniform distribution of the particle positions over the
whole slice while the microstructure on the right presents locally
more densely packed particles and consequently larger voids. Both
structures have an identical overall volume fraction of 0.4.

In preceding works, various characterization methods, such as the
radial pair correlation function~\cite{Huetter_2000}, the bond
angle distribution function~\cite{Huetter_2000}, the triangle
distribution function~\cite{Huetter_2000} and the Minkowski
functionals in conjunction with the parallel-body
technique~\cite{Huetter_2003} were applied to sets of
microstructures generated by BD simulations~\cite{Huetter_1999}.

The pair correlation function quantifies the amount of structural
rearrangement during the coagulation. Its usefulness regarding a
distinction between structures with different DOH however is
rather limited as the differences between peaks corresponding to
characteristic particle separation distances are relatively
small~\cite{Huetter_2000}. The main advantage of the pair
correlation function is its experimental accessibility through
scattering techniques such as SESANS~\cite{Andersson_2008}.

The bond angle distribution function and the triangle distribution
function give further information on the local building blocks of
the particle network~\cite{Huetter_2000}. Particular features, as
for example peaks in the respective distribution function allow
distinguishing between "more homogeneous" and "more heterogeneous"
microstructures. However, as in the case of the pair correlation
function, the differences between structures with different DOH
are small for both, the bond angle and the triangle distribution
function.

The analysis using the Minkowski functionals in conjunction with
the parallel body technique supplies additional information on the
structure's morphology resolving microstructural differences on a
length scale limited by the largest pore size~\cite{Huetter_2003}.
This method is computationally intensive and the extension to
arbitrary particle shapes is difficult.

Gearing towards a possible correlation between microstructure and
mechanical properties "homogeneous" and "heterogeneous"
microstructures have recently been analyzed in terms of load
bearing sub-structures: Firstly, regions of closely packed
particles and secondly, quasi-linear chains of contacting
particles~\cite{Schenker_2008}. The locally closed packed regions
were analyzed using the common neighbor distribution in
conjunction with the dihedral angle distribution. Both methods
only showed minor differences between a "homogeneous" and a
"heterogeneous" microstructure. In particular, practically the
same number of triangles and regular tetrahedrons were found in
both structures. Quasi-linear arrangements of contacting particles
were quantified using the straight path method, revealing
significant microstructural differences between "homogeneous" and
"heterogeneous": approximately twice as many paths of length
longer or equal to four particles and three times as many paths of
length longer or equal to five particles where found in the
"heterogeneous" microstructure.

Despite the multitude of microstructural characterization methods
that have been applied to colloidal microstructures possessing
various DOH, a useful quantification of the microstructures'
heterogeneity in form of a scalar measure is lacking. In the
present study statistical measures allowing for a quantification
of a structure's heterogeneity are provided. The following three
methods aiming at this quantification are discussed.

Firstly, the exclusion probability~\cite{Torquato_1990} estimates
the pore size distribution by randomly probing the pore space.
In~\cite{Whittle_1999} this method was applied to very dilute
simulated colloidal systems and allowed for a clear distinction
between particle gel networks with varying textures. The same
method, termed spherical contact distribution
function~\cite{Bezrukov_2002}, was used to investigate the pore
size distribution of dense sphere packings as a function of the
particle size distribution and the packing generation algorithm.

Secondly, the density fluctuation method considers the fixed
particle centers as a point process, and it statistically analyses
the fluctuations of the particle center density as a function of
length scale. Used comparatively, this method further allows for a
detailed analysis of the length scale, on which two structures
present the largest differences in terms of heterogeneity.

Finally, the Voronoi volume distribution~\cite{Voronoi_1908} is
used to quantify the distribution of the free volume of our
particle packings. In~\cite{Yang_2002} the packing of cohesive
particles resulting from simulations using the discrete element
method~\cite{Cundall_1979} with volume fractions between
approximately 0.2 and 0.6 was analyzed. The distribution of
Voronoi volumes was shown to broaden with decreasing volume
fraction. In~\cite{Aste_2008} the Voronoi volume distribution was
determined for a large set of experimental and numerical data
covering a wide range of volume fractions. The various
distributions were shown to follow a so-called $k$-gamma function,
which was deduced by means of a statistical mechanics approach.
The parameter $k$ characterizing the shape of the curve was found
to depend very sensitively on the structural organization of the
particles.

To the authors' knowledge, none of these methods have yet been
applied to particle structures, for which the DOH is the only
variable, in opposition to a varying volume fraction or particle
size distribution for example.

\section{\label{sec:MatAndMeth}Materials and Methods}

In the following, the structure characterization methods employed
in this work are presented: the pore size distribution, the
density fluctuation method and the distribution of Voronoi
volumes. These methods are evaluated in terms of their ability to
quantify the DOH of microstructures generated by previous BD
simulations~\cite{Huetter_2000,Huetter_1999}. In these
simulations, the DLVO-theory~\cite{Russel_1989} was used to
describe the inter-particle potential $V^{dlvo}$ given by the sum
of an attractive van der Waals term $V^{vdw}$ (Eq.~\ref{eq:Vvdw})
and an electrostatic repulsion term $V^{el}$
(Eq.~\ref{eq:Vel}).Thus, $V^{dlvo} = V^{vdw} + V^{el}$ with,

 \begin{equation}
    V^{vdw}(r) =  \frac{-A_H}{12}\left[\frac{d_0^2}{r^2-d_0^2} + \frac{d_0^2}{r^2} + 2 \ln \left(\frac{r^2-d_0^2}{r^2}\right)\right]
 \label{eq:Vvdw}
 \end{equation}

and

\begin{eqnarray}
 V^{el}(r) =&&  \pi \epsilon_r \epsilon_{\mathrm{0}}
    \left[ \frac{4 k_b T}{z e} \tanh \left( \frac{z e}{4 k_b T} \Psi_{\mathrm{0}} \right)
    \right]^2 \nonumber\\
&&\times d_0 \exp \left( -\kappa \{ r-d_0\} \right) \label{eq:Vel}
\end{eqnarray}

respectively. The DLVO parameters are summarized in
Table~\ref{tab:BDparameter}.

\begin{table}
\caption{\label{tab:BDparameter} DLVO Potential Parameters.}
\begin{ruledtabular}
\begin{tabular}{lll}
Parameter & Symbol & Value  \\
\hline Hamaker constant of Al$_2$O$_3$ in H$_2$O & $A_H$ & $4.76 \times 10^{-20}$ J\\
Particle diameter  & $d_0$ & 5 $\times$ 10$^{-7}$ m \\
Relative dielectric constant of H$_2$O & $\epsilon_{r}$ & 81 \\
Surface potential & $\Psi_0$ & 0 - 15 mV\\
Absolute temperature  & $T$ & 293 K \\
Valency of ions  & $z$ & 1 \\
Inverse Debye screening length & $\kappa$ & 10$^{8}$ m$^{-1}$ \\
\end{tabular}
\end{ruledtabular}
\end{table}

The heterogeneity of the final microstructure was shown to be
closely related to the presence and depth of the secondary minimum
in the inter-particle potential, which, for fixed values of the
Debye screening length, essentially depends on the particles'
surface potential $\Psi_0$~\cite{Wyss_1_2005,Huetter_2000}.

The microstructures analyzed in this work are labelled according
to the surface potential $\Psi_0$ used during their generation. In
particular, the following $\Psi_0$-values are used: 0 mV, 12 mV,
13 mV, 14 mV and 15 mV. Additionally, these final microstructures
are compared to the initial microstructure, representing a
stabilized suspension in which the inter-particle potential is
purely repulsive. The volume fraction is fixed at 0.4, the
monosized particles have a radius $r_0$ = 0.25 $\mu$m, all
microstructures consist of 8000 particles and are contained in a
simulation box with periodic boundary conditions. In particular,
the particle interpenetration is much smaller than the length
scale of the heterogeneities analyzed in this study. Less than
0.1\% of all contacts present an interpenetration above 1.0\% of
the particle diameter $d_0$ with a maximum interpenetration of
1.3\% $d_0$. Please refer to~\cite{Huetter_2000,Huetter_1999} for
a more detailed description of the BD-simulations.

\subsection{\label{sec:psd1}Pore Size Distribution}

\begin{figure}
\includegraphics[width=\linewidth]{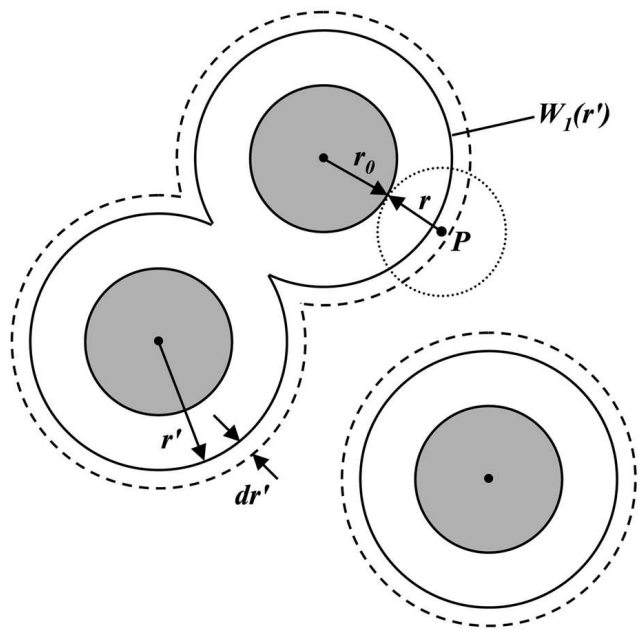}
\caption{\label{fig:fig_2} Particle structure (gray particles)
with a test particle inserted at position $P$.}
\end{figure}

The pore size distribution is estimated following the approach
described by Torquato et al.~\cite{Torquato_1990} using the
exclusion probability $E_V(r)$. $E_V(r)$ is defined as the
probability of inserting a "test" particle of radius $r$ at some
arbitrary position in the pore space of a system of $N$ particles.
This is schematically represented in Fig.~\ref{fig:fig_2} using a
set of particles of radius $r_0$ (gray) with a test particle of
radius $r$ inserted at position $P$.

In order to estimate $E_V(r)$, a statistically large number of
points is randomly placed in the pore space of a given
microstructure and the distance to the closest particle surface is
determined.

In~\cite{Huetter_2003} the relation between $E_V(r)$ and the
Minkowski functional $W_1(r)$ was described. The Minkowski
functional in conjunction with the parallel body technique
considers the point process given by the fixed particle centers.
Generally, in three dimensions there are four functionals $W_i$,
where $i = 0, .., d$ with $d$ the spatial dimension, corresponding
to the volume, surface, average mean curvatures and connectivity.
In particular, $W_1(r') = \frac{1}{3}\int_{\partial A(r')}dS$
calculates the surface of the union of spheres located at the
particle centers in dependence of their radius $r' = r_0 + r$.
Schematically, $W_1(r')$ is shown in Fig.~\ref{fig:fig_2}.
$W_1(r')dr'$ is the volume between the distance $r'$ and $r' +
dr'$. The probability of placing an uniformly drawn random test
point at a distance $r \in [r', r' + dr'[$ is proportional to the
volume delimited by $r'$ and $r' + dr'$ and therefore $P(r' \leq
r_0 + r < r' + dr') = E_V(r)dr' \propto W_1(r')dr'$, which results
in $E_V(r_0+r) \propto W_1(r')$. Thus, given a statistically large
number of test points $E_V(r)$ provides a means to estimate
$W_1(r')$, which has the advantage of being computationally less
intensive. A similar Monte Carlo integration is usually performed
to calculate $W_0(r')$.

\subsection{\label{sec:df1}Density Fluctuations}

The density fluctuation method quantifies the local fluctuation of
particle centers by subdividing the structure into smaller parts
and measuring the average value and the standard deviation of the
particle center density. Practically, this is done by dividing the
structure into $n_c^3$ cells using a cubic grid, where $n_c$ is
the number of cells along one dimension with $n_c$ = 2, ..,
$n_c^{max}$, $n_c^{max}$ being the maximum number of cells under
consideration (along one dimension). The density fluctuation
method consists in determining the average number of particle
centers per cell $E_{ppc}$ and its standard deviation
$\sigma_{ppc}$ as a function of $n_c$ and then calculating the
relative fluctuations $\frac{\sigma_{ppc}}{E_{ppc}}$.

Additionally, the density fluctuation method is applied
comparatively allowing for a determination of the length scale on
which two structures show the largest differences. Therefore, the
difference $\Delta(n_c)$ given in Eq.~(\ref{eq:one}) is calculated
between two structures $i$ and $j$.

\begin{equation}
\Delta(n_c) = \left[ \frac{\sigma_{ppc}^i}{E_{ppc}^i} -
\frac{\sigma_{ppc}^j}{E_{ppc}^j} \right](n_c)
 \label{eq:one}
\end{equation}

If both structures have the same number of particles and identical
volume fractions, which is the case for the microstructures
investigated in this work, then $E_{ppc}^i = E_{ppc}^j = E_{ppc}$
and Eq.~(\ref{eq:one}) yields Eq.~(\ref{eq:two}).

\begin{equation}
\Delta(n_c) = \frac{\sigma_{ppc}^i - \sigma_{ppc}^j}{E_{ppc}}(n_c)
 \label{eq:two}
\end{equation}

$\Delta$ is plotted against the cell's edge length $l_c(n_c) =
L_{Box}/n_c$ normalized by the particle diameter $d_0$, where
$L_{Box}$ is the side length of the cubic simulation box.

\subsection{\label{sec:vvd1}Voronoi Volume Distribution}

Formally, for a set of monodispersed spherical particles, the
Voronoi volume $V_i$ associated with a particle $i$ is a
polyhedron whose interior consists of all points in space that are
closer to the center of particle $i$ than to any other particle
center~\cite{Voronoi_1908}. The Voronoi tessellation thus divides
the volume containing a set of particles into a set of
space-filling, non-overlapping and convex polyhedrons. In this
work, the Quickhull algorithm~\cite{Barber_1996} is used to
compute the volumes of the Voronoi polyhedrons. The distribution
of the Voronoi volumes describes the deviation of a structure from
a perfect crystalline packing, in which case all particles occupy
the same volume and the Voronoi volume distribution thus is a
delta function. The minimum volume of a Voronoi cell $V_{min}$ is
achieved for a regular close packing with $V_{min} = 1.325
V_{sphere}$, where $V_{sphere}$ is the volume of a particle. The
difference between a particle's Voronoi volume $V_i$ and $V_{min}$
is termed the Voronoi free volume $V_i^f = V_i - V_{min}$.

The distribution of the Voronoi free volume was found to follow
gamma-distributions: Kumar and Kumaran for example have shown that
the free volume distribution of hard-disk and hard-sphere systems
are well described using a two-parameter and a three-parameter
gamma-distributions~\cite{Kumar_2005}.

Aste et al. have deduced the two-parameter gamma-distribution
using a statistical mechanics approach~\cite{Aste_2008}. The
so-called $k$-gamma distribution given by Eq.~(\ref{eq:three}) was
found to agree very well with a large number of experiments and
computer simulations over a wide range of packing fractions.

\begin{equation}
f(V^f, k) = \frac{k^k}{\Gamma(k)}
\frac{(V^f)^{k-1}}{(\bar{V}^f)^k}\exp(-k \frac{V^f}{\bar{V}^f})
 \label{eq:three}
\end{equation}

The mean Voronoi free volume $\bar{V}^f$ is a scaling parameter
given by $\bar{V} - V_{min} = \left( 1 / \Phi - 1.325 \right)
V_{sphere}$, where $\Phi$ is the volume fraction. The free
parameter $k$ characterizing the shape of the curve depends very
sensitively on the structural organization of the particles and
corresponds to the specific heat in classical thermodynamics.
Empirically, $k$ can be computed using $k =
\frac{(\bar{V}^f)^2}{\sigma_V^2}$, where $\sigma_V^2$ is the
variance of the free volume distribution. In particular,
Eq.~(\ref{eq:three}) was shown to hold for systems at statistical
equilibrium as well as for systems out of
equilibrium~\cite{Aste_2008}.

\section{\label{sec:ResAndDisc}Results and Discussion}

\subsection{\label{sec:psd2}Pore Size Distribution}

\begin{figure}
\includegraphics[width=\linewidth]{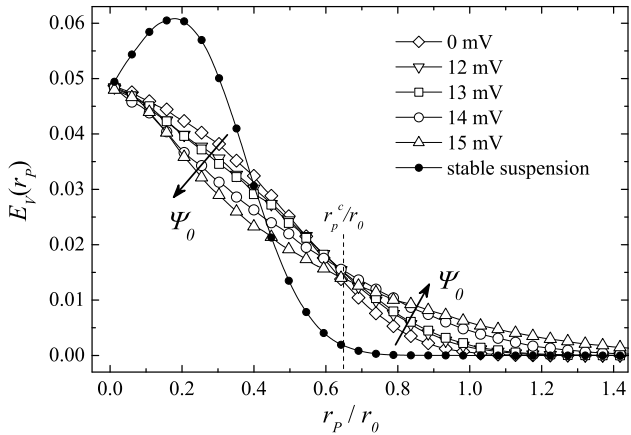}
\caption{\label{fig:fig_3} Pore size distribution for the initial
microstructure (stable suspension) and final microstructures
($\Psi_0$ = 0 - 15 mV). B-spline curves serve as guide to the
eye.}
\end{figure}

The pore size distribution of the various microstructures is
depicted in Fig.~\ref{fig:fig_3} in terms of the exclusion
probability $E_V(r_P)$ as a function of the pore radius $r_P$
normalized by the particle radius $r_0$. A set of $10^6$ random
test points~\cite{Matsumoto_1998} placed in the structures' pore
space was used to estimate $E_V(r_P)$.

All curves for the final microstructures ($\Psi_0$ = 0 - 15 mV)
decrease monotonically towards increasing pore sizes indicating a
decreasing probability of finding larger pores. A particular
behavior is found for the stable suspension where the exclusion
probability increases with increasing pore size up to a pore
radius of $0.18 r_0$. This is due to the repulsive inter-particle
potential in the case of the stable suspension where consequently,
the particles are not in contact. The pore diameter, at which the
maximum in the exclusion probability is found, corresponds to the
average surface-to-surface distance between neighboring particles.

Remarkably, the various curves for the final microstructures in
Fig.~\ref{fig:fig_3} intersect at approximately the same point
defining a characteristic pore size $r_p^c$, found at $0.65 r_0$.
The probability of finding a pore with a radius below $r_p^c$
decreases for increasing values of $\Psi_0$ while pores with a
radius above $r_p^c$ are found with higher probability towards
increasing $\Psi_0$. Indeed, the probability of finding pore radii
larger than approximately $1.1 r_0$ is negligible in the case of
the "most homogeneous" microstructure with $\Psi_0$ = 0 mV while
pore radii up to $2.4 r_0$ are found in the "most heterogeneous"
microstructure with $\Psi_0$ = 15 mV.

\begin{figure}
\includegraphics[width=\linewidth]{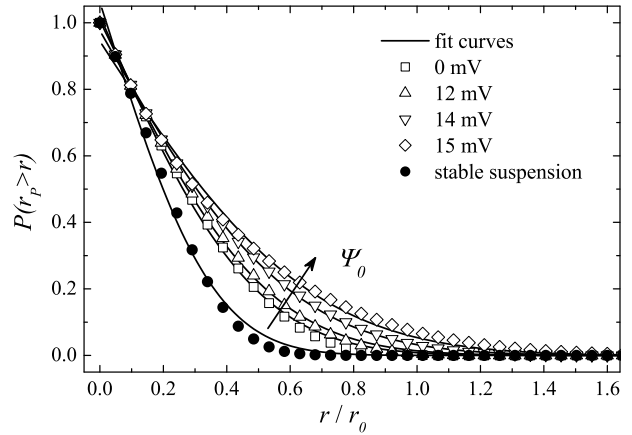}
\caption{\label{fig:fig_4} Probability of finding pores with a
radius $r_P$ larger than $r$ vs. $r$ normalized by the particle
radius $r_0$ for the various microstructures (symbols) and
corresponding fits using a complementary error function (solid
lines).}
\end{figure}

Using the results obtained for the exclusion probability
$E_V(r_P)$, the cumulative probability $P(r_P > r)$ of finding
pore radii larger than $r$ was calculated using $P(r_P > r) =
\sum_{r' > r_P}E_V(r')$. The results are shown in
Fig.~\ref{fig:fig_4}.

$P(r_P > r)$ decreases monotonically for all microstructures. The
fastest decrease is found for the stable suspension. With
increasing $\Psi_0$ the decrease of $P(r_P > r)$ is slower.
Comparing the "most and the least heterogeneous" microstructure,
with $\Psi_0$ = 15 mV and $\Psi_0$ = 0 mV, respectively, there is
a 1.7 times higher probability of finding pores larger than $0.5
r_0$. Towards larger pore radii, the probability ratio increases:
finding pores with a radius larger than $0.75 r_0$ and $1.0 r_0$
is 5.1 times and 60 times, respectively, more probable in the
"heterogeneous" than in the "homogeneous" microstructure.
Fig.~\ref{fig:fig_4} further shows the fit of $P(r_P > r)$ using a
complementary error function given by Eq.~(\ref{eq:four}).

\begin{equation}
P(r_P > r) = 1 - \mathrm{erf} \left( \frac{r/r_0 -b}{a \sqrt{2}}
\right)
 \label{eq:four}
\end{equation}

The error function is defined as the cumulative Gaussian
distribution: $\mathrm{erf}(x) =2/\sqrt{\pi} \int_0^x \exp \left(
-  z^2  \right) dz$. Parameter $a$ is the standard deviation, i.e.
the width of the corresponding Gaussian distribution and $b$ is
the location of its maximum, i.e. the most probable pore to
particle radius ratio.

Table~\ref{tab:table1} summarizes the fit parameter $a$ and $b$
obtained for the various microstructures analyzed in this work and
the corresponding correlation coefficients R$^2$, which for all
fits are very close to 1 and thus indicate a good fit. Parameter
$a$ is smallest for the initial microstructure and increases
towards increasing values of $\Psi_0$. The increasing values of
$a$ reflect the slower decrease of the curves in
Fig.~\ref{fig:fig_4} and hence the broadening of the distributions
towards increasing DOH. The values found for $b$ decrease with
increasing $\Psi_0$ representing a shift of the maximum in the
Gaussian distribution shown in Fig.~\ref{fig:fig_3}.

In~\cite{Torquato_1990} the following expression for $E_V(r)$ was
found for a statistically homogeneous microstructure of
impenetrable spheres: $E_V(r) = (1-\Phi) \exp(P_3(r,\Phi))$, where
$\Phi$ is the volume fraction and $P_3$ is a third degree
polynomial function in $r$. This function can be interpreted as a
corrected Gaussian distribution which is nicely approximated by
Eq.~(\ref{eq:four}) as well (R$^2 = 0.9962$), yielding $a = 0.347$
and $b = 0.0204$. The DOH of this theoretical structure thus lies
between the stable suspension and the most homogeneous final
microstructure with $\Psi = 0$ mV.

An alternative to the quantification of a structure's
heterogeneity by means of the fit parameter $a$ is the calculation
of the integral over the cumulative pore size distribution. This
scalar measure has the advantage of being statistically more
robust. It is also more general in the sense that it is applicable
even when the fit with a complementary error function does not
yield good results. The integral over the cumulative pore size
distribution is labelled $I_{ps}$ and is given by
Eq.~(\ref{eq:seven}):

\begin{equation}
I_{ps} = \int_{r > 0} P(r_P > r) \frac{dr}{r_0} = \frac{\delta
r}{r_0} \sum_{r_i > 0} P(r_P > r_i)
 \label{eq:seven}
\end{equation}

The second equality accounts for the discrete case, where $\delta
r$ is the radial resolution of the empirical pore size
distribution. The $I_{ps}$-values for the various microstructures
are summarized in Table~\ref{tab:table1}. In our case, in which
the data can nicely be fitted using Eq.~(\ref{eq:four}), the fit
parameter $a$ is proportional to $I_{ps}$: $a/I_{ps} = 1.31 \pm
0.05$. Thus, a quantification of the DOH by means of $a$ or
$I_{ps}$ is equivalent.

\begin{table}
\caption{\label{tab:table1}Fit parameters $a$ and $b$,
R$^2$-values and integrals $I_{ps}$ and $I_{df}$ for the various
microstructures.}
\begin{ruledtabular}
\begin{tabular}{lccccc}
 & $a$ & $b$ ($10^{-2}$) & R$^2$
 & $I_{ps}$ & $I_{df}$ \\
\hline stable suspension& 0.2651 & 2.042 & 0.9952
& 0.215 & 20.29 \\
$\Psi_0$ = 0 mV & 0.3752 & 0.9013 & 0.9987
& 0.291 & 22.01 \\
$\Psi_0$ = 12 mV & 0.4053 & 0.3136 & 0.9994
& 0.311 & 22.31 \\
$\Psi_0$ = 13 mV & 0.4127 & 0.1225 & 0.9996
& 0.315 & 22.44 \\
$\Psi_0$ = 14 mV & 0.4736 & -1.415 & 0.9996
& 0.352 & 23.05 \\
$\Psi_0$ = 15 mV & 0.5377 & -3.756 & 0.9972
& 0.388 & 23.71 \\
\end{tabular}
\end{ruledtabular}
\end{table}

\subsection{\label{sec:df2}Density Fluctuations}

\begin{figure}
\includegraphics[width=\linewidth]{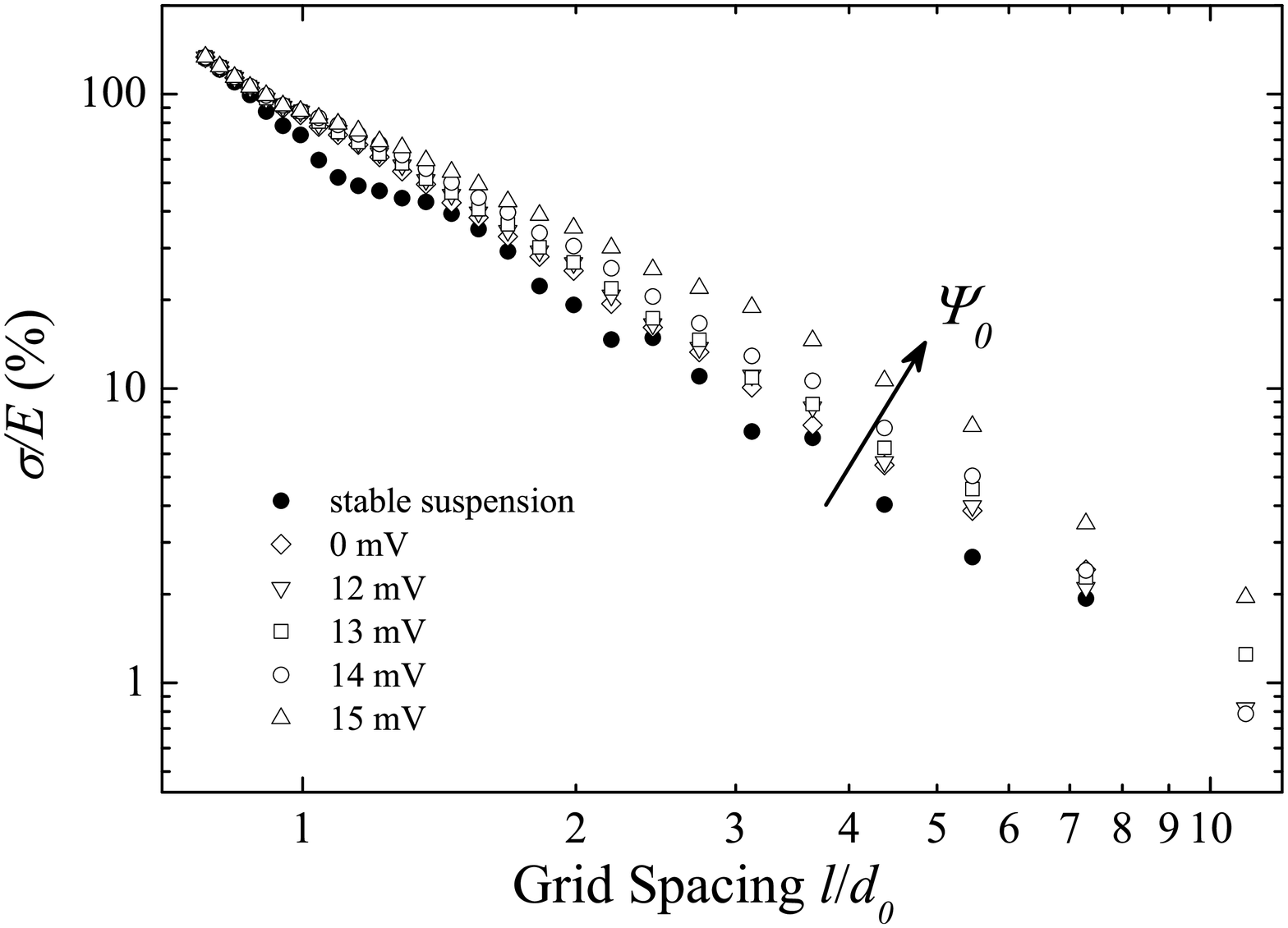}
\caption{\label{fig:fig_5} Relative density fluctuations for the
various final microstructures and the stable suspension as a
function of grid spacing.}
\end{figure}

The density fluctuations for the various microstructures are shown
in Fig.~\ref{fig:fig_5}. Over the whole range of grid spacings,
the fluctuations are smallest for the stable suspension and
increase for increasing values $\Psi_0$. For $n_c \geq 34$, which
corresponds to a grid spacing of 0.64 particle diameter, the
density fluctuations of the various microstructures are equal.
$I_{df}$ given in Eq.~(\ref{eq:eight}) provides an integral
measure of the heterogeneity similar to $I_{ps}$ in the previous
section, however accounting for the discrete variable $n_c$.

\begin{equation}
I_{df} = \sum_{n_c < 34} \frac{\sigma_{ppc}}{E_{ppc}}(n_c)
 \label{eq:eight}
\end{equation}

The $I_{df}$ values for the various microstructures summarized in
Table~\ref{tab:table1} continuously increase towards increasing
values of $\Psi_0$ and thus measure the DOH of the
microstructures.

In the following, two sets of comparisons are performed: Firstly,
the various final microstructures are compared to the stable
suspension. This comparison quantifies the length scale on which
structural rearrangements take place during the coagulation.
Secondly, the final microstructures with $\Psi_0 > 0$ mV are
compared to the "most homogeneous" microstructure with $\Psi_0$ =
0 mV. This set of comparisons reveals the length scale on which
variations in heterogeneity are most pronounced.

\begin{figure}
\includegraphics[width=\linewidth]{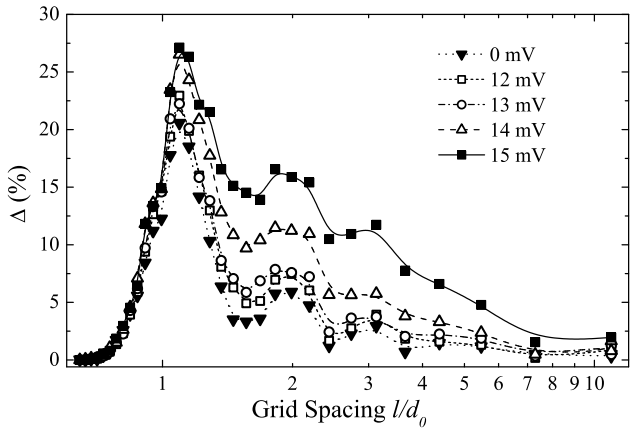}
\caption{\label{fig:fig_6} Relative difference between the density
fluctuations of the various final microstructures and the stable
suspension as a function of grid spacing. B-spline curves serve as
guide to the eye.}
\end{figure}

The comparison of the various final microstructures to the
initial, stabilized microstructure is shown in
Fig.~\ref{fig:fig_6} in terms of $\Delta(n_c)$ as given by
Eq.~(\ref{eq:two}) where superscripts $i$ and $j$ correspond to a
final microstructure and to the initial microstructure,
respectively. $\Delta(n_c)$ is shown as function of the grid
spacing $l$ normalized by the particle diameter $d_0 = 2r_0$.

All curves present an identical behavior, which essentially
consists of three successive peaks with decreasing height towards
a larger grid spacing. The location of the first, second and third
peak is slightly above one, at two and at three particle
diameters, respectively. The height of the individual peaks
increases for increasing values $\Psi_0$.

More precisely speaking, the first peak is found at $1.09 d_0$ for
all final microstructures in comparison to the stable suspension.
This grid spacing corresponds to a cell number of 8000 and
therefore to the case where the average number of particles per
cell is exactly one. This case is best reproduced for the stable
suspension as for the final microstructures the standard deviation
is roughly 20 to 27\% larger. Physically, this peak is explained
by the transition of the inter-particle potential from repulsive
to attractive. Indeed, the repulsive potential in the case of the
stable suspension causes all particles to occupy approximately the
same volume as will be confirmed in Sec.~\ref{sec:vvd2} by means
of the Voronoi volume distribution. The switching of the
inter-particle potential from repulsive to attractive causes the
particles to form contacts resulting in an average particle
separation of one particle diameter, which is smaller than the
grid spacing of $1.09 d_0$. This increases the probability of
finding cells that are either empty or contain more than one
particle and thus the standard deviation of the average number of
particles per cell is increased.

The peaks at a grid spacing of approximately two and three
particle diameters are considerably less pronounced than the peak
at $1.09 d_0$. In particular, the differences between the various
final microstructures are larger than for the first peak. These
differences will be elaborated in more detail in the following.

\begin{figure}
\includegraphics[width=\linewidth]{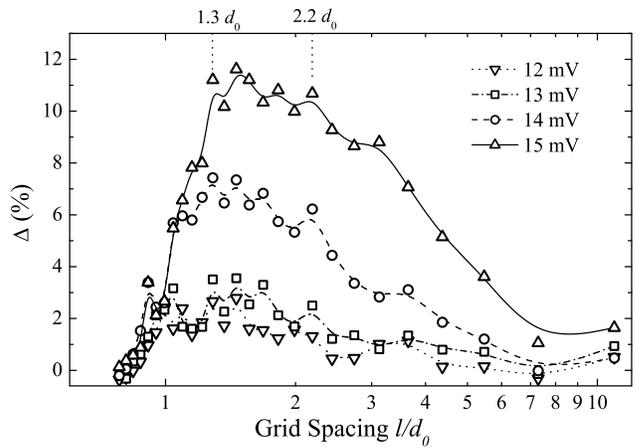}
\caption{\label{fig:fig_7} Relative difference between the density
fluctuations of the various microstructures with ($\Psi_0 >$ 0 mV)
and the "most homogeneous" microstructure ($\Psi_0$ = 0 mV) as a
function of grid spacing. B-spline curves serve as guide to the
eye.}
\end{figure}

The comparison between the final microstructures with $\Psi_0 > 0$
mV and the "most homogeneous" microstructure with $\Psi_0$ = 0 mV
is shown in Fig.~\ref{fig:fig_7}. Here, superscripts $i$ and $j$
(Eq.~(\ref{eq:two})) correspond to one of the microstructures with
$\Psi_0 >$ 0 mV and to the microstructure with $\Psi_0$ = 0 mV,
respectively. Over the whole range of grid spacings, the
differences between the density fluctuations increase towards
higher values of $\Psi_0$. This behavior correlates very well with
the increase of porosity for increasing $\Psi_0$ as already
observed in the previous section. Additionally,
Fig.~\ref{fig:fig_7} reveals that the largest differences in terms
or particle density between the "most and least heterogeneous"
microstructure ($\Psi_0$ = 15 mV and $\Psi_0$ = 0 mV,
respectively) are found on a length scale between 1.3 and 2.2
particle diameters.

\subsection{\label{sec:vvd2}Voronoi Volume Distribution}

As stated in Sec.~\ref{sec:vvd1} the distribution of Voronoi
volumes $P(\nu)$ describes the deviation of a given structure from
a perfectly crystalline packing, for which $P(\nu)$ is a delta
function and all particles occupy the same volume. For random
particle structures $P(\nu)$ broadens and as will be shown in the
following, the width of the distribution can be interpreted as the
heterogeneity of a structure.

\begin{figure}
\includegraphics[width=\linewidth]{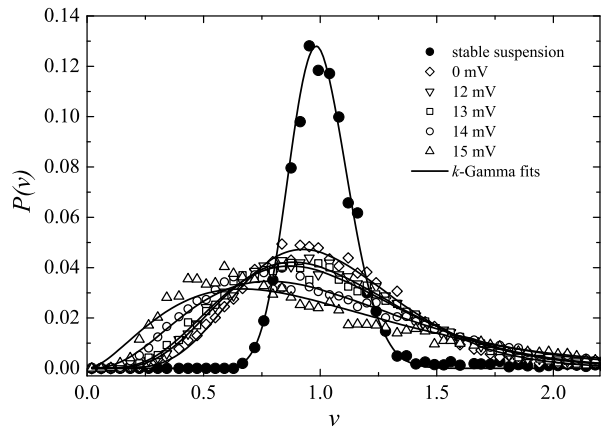}
\caption{\label{fig:fig_8} Voronoi volume distribution $P(\nu)$
 for the various microstructures (symbols) and corresponding fits using
 the $k$-gamma distribution (lines).}
\end{figure}

We've calculated $P(\nu)$ for the stable suspension and the
various final microstructures as a function of $\nu = \frac{V -
V_{min}}{\bar{V} - V_{min}}$, the free volume normalized by the
mean free volume (Fig.~\ref{fig:fig_8}, symbols). The distribution
found for the stable suspension is significantly narrower in
comparison to the final microstructures. In the case of the final
microstructures, $P(\nu)$ broadens with increasing value of
$\Psi_0$ indicating that larger fluctuations in Voronoi volumes
are found with increasing heterogeneity.

The various microstructures were fitted using the $k$-Gamma
distribution given in Eq.~(\ref{eq:three}). The resulting curves
are shown in Fig.~\ref{fig:fig_8} (lines) and the corresponding
$k$- and R$^2$-values are summarized in Table~\ref{tab:table2}.
$k$ decreases with increasing heterogeneity and can therefore be
used as a measure for the DOH of the microstructures. The
R$^2$-values close to one indicate good fits. In particular, a
very good fit quality was achieved for the stable and the final
microstructures up to $\Psi_0$ = 14 mV. The R$^2$-value for the
$\Psi_0$ = 15 mV microstructure is lower. Indeed,
Fig.~\ref{fig:fig_8} shows that for the $\Psi_0$ = 15 mV
microstructure the agreement between the measured distribution and
the fit curve for small Voronoi volumes is not as good as for the
other curves. This might be related to the fact that during the
generation of the $\Psi_0$ = 15 mV microstructure the energy
barrier between primary and secondary minimum in the
inter-particle potential was largest. This resulted in a few
particle contacts still trapped in the secondary minimum (roughly
7\% of the physical contacts). Particles trapped in the secondary
minimum have an inter-particle distance of $2.16 r_0$ instead of
$2 r_0$ upon complete coagulation, which may be a reason for the
reduced fit quality towards smaller Voronoi volumes.

\begin{table}
\caption{\label{tab:table2}$k$-Gamma fit results.}
\begin{ruledtabular}
\begin{tabular}{lcc}
 & $k$ & R$^2$ \\
\hline stable suspension& 62.3 & 0.990 \\
$\Psi_0$ = 0 mV & 8.6 & 0.995 \\
$\Psi_0$ = 12 mV & 6.5 & 0.996 \\
$\Psi_0$ = 13 mV & 6.0 & 0.996 \\
$\Psi_0$ = 14 mV & 4.0 & 0.994 \\
$\Psi_0$ = 15 mV & 2.8 & 0.972 \\
\end{tabular}
\end{ruledtabular}
\end{table}

\section{\label{sec:SumAndConc}Summary and Conclusions}

In this paper, we have analyzed three distinct microstructural
characterization methods. Using these methods, scalar measures
were introduced, which for the first time allow quantifying the
DOH of particle packings.

\begin{itemize}
 \item{
The exclusion probability gives an estimate of the pore size
distribution by a random probing of the pore space. The "more
heterogeneous" microstructures present a considerably broader pore
size distribution with a significantly longer tail than the
distribution for the "more homogeneous" microstructures. In
particular, a continuous broadening is found with increasing
heterogeneity. The cumulative exclusion probabilities were shown
to follow error functions with parameter $a$ reflecting their
width and thus measuring the structures' DOH. Fit parameter $a$
increases with increasing heterogeneity.}

 \item{
The density fluctuation method statistically analyzes the particle
center density in dependence of the sampling length scale. The
relative density fluctuation as function of the grid spacing
presents a clear dependence on the heterogeneity of the
microstructure: Over the whole range of grid spacings, the stable
suspension exhibits the smallest density fluctuations. These
fluctuations increase towards increasing values of $\Psi_0$ and
thus increasing DOH, which is nicely reflected by increasing
values of the integral measure $I_{df}$.

An examination of the differences between the density fluctuations
of two structures was found to be particularly useful as it allows
determining the length scale on which the structures present the
largest differences in heterogeneity. In the case of the "most
heterogeneous" microstructure the largest differences in
comparison to the "most homogeneous" one are found on a length
scale between 1.3 and 2.2 particle diameters.}

 \item{
The Voronoi volume distribution of the stable suspension is very
narrow in comparison to the final microstructures, for which the
distribution broadens with increasing heterogeneity. The various
Voronoi volume curves were shown to follow $k$-gamma
distributions. Parameter $k$, reflecting the width of the
distribution and thus the structure's DOH, decreases with
increasing heterogeneity.}
\end{itemize}

\begin{figure}
\includegraphics[width=\linewidth]{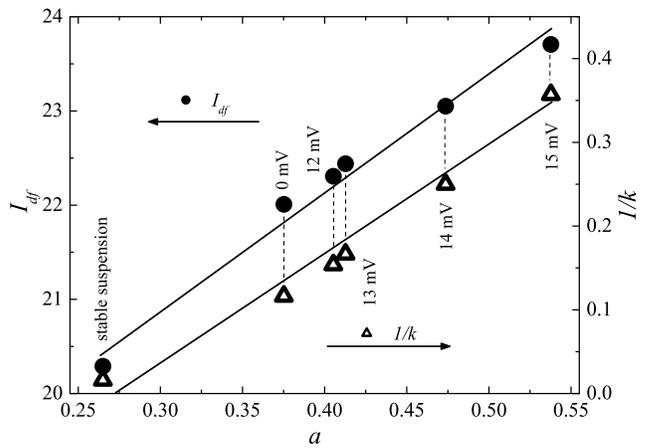}
\caption{\label{fig:fig_9} Interdependence between the measures of
the degree of heterogeneity for the various microstructures:
$I_{df}$ and $1/k$ as a function of $a$.}
\end{figure}

The behavior of the three parameters $a$, $k$ and $I_{df}$ is
summarized in Fig.~\ref{fig:fig_9} showing $I_{df}$ (left scale)
and $1/k$ (right scale) as a function of $a$. The solid lines
suggest a pairwise affine dependence between $I_{df}$, $1/k$ and
$a$. Thus, as far as the quantitative characterization of the DOH
of the microstructures considered in this work is concerned, all
methods, the pore size distribution, the density fluctuation
method and the Voronoi volume distribution, can be considered as
equivalent in the sense that the knowledge of one parameter
permits to determine the others. However, parameter $k$ reflects
changes in the DOH more sensitively than $a$ or $I_{df}$. Indeed,
the normalization of $k$, $a$ and $I_{df}$ with respect to their
maximum values reveals that parameter $k$ covers the interval
$[0.04, 1.0]$. This interval is significantly larger than the
normalized ranges of $a$ and $I_{df}$, which are $[0.5, 1.0]$ and
$[0.86, 1.0]$, respectively.

The interrelation between the three structural characterization
methods can be understood as follows: The probability of placing a
random point used for the determination of the pore size
distribution into the free Voronoi volume of a particle is
proportional to the particle's free Voronoi volume. Thus, the
broader the distribution of Voronoi volumes, the higher the
probability of finding larger pores, which leads to a longer tail
in the exclusion probability as shown in Fig.~\ref{fig:fig_3}. The
relation between the Voronoi volume distribution and the density
fluctuation follows similar arguments: A broadening in the Voronoi
volume distribution essentially means that there is a broader
distribution in the nearest neighbor distances and therefore
larger differences in the density fluctuations.

We have applied the methods to a set of monodispersed spherical
particle packings representing stable and coagulated colloidal
particle structures, but the methods could of course be
generalized. The pore size distribution as determined by the Monte
Carlo method employed in this work can be applied as it is to any
porous media. In this sense, it is the most general method
analyzed in this study. The fit using an error function however
may not necessarily yield good results. In this case the integral
measure $I_{ps}$ proposed in Sec.~\ref{sec:psd2} could be used or
the width of the distribution could be determined empirically. The
Voronoi volume distribution generally only requires that the
elements constituting a structure are convex and in this case, the
empirical distribution can be determined. To the authors'
knowledge a fit using the $k$-gamma distribution has however only
been performed in the case of packings of monodispersed, spherical
particles. As for the density fluctuation method, we have in this
work considered the density of the particle centers. The method
may be extended to a determination of the exact portion of the
sphere volumes per cell, which however is computationally
expensive. An alternative could be a cell-wise Monte Carlo
integration of the partial sphere volumes, which would allow for a
characterization of arbitrary porous structures using the density
fluctuation method.

From an experimental point of view, the methods presented in this
study rely on the possibility to determine the particle positions,
which in the case of colloidal particles can be obtained using
confocal laser microscopy for example~\cite{Crocker_1996}. In
particular, the pore size distributions measured using mercury
porosimetry~\cite{Giesche_2006} and estimated using the exclusion
probability are not equivalent since the latter overestimates the
number of small pores due to the random probing of the pore space.

In this paper, we have introduced three scalar measures, which for
the first time allow quantifying and comparing the heterogeneity
of packings of spherical particles in terms of a DOH. These
measures were calculated using distinct techniques and structural
characterization methods. In view of these differences, the very
nice correlation between the three DOH-measures is remarkable.
Indeed, it suggests that the DOH is a microstructure's inherent
property and that any of the methods proposed in this work can be
used to uniquely characterize and classify it. In terms of
sensitivity however, considerable differences between the methods
where found. Parameter $k$ reflects differences in the DOH most
sensitively, followed by parameter $a$ and finally $I_{df}$. A
further definition of an absolute DOH would require a suitable
reference structure, which for example is either perfectly
heterogeneous or perfectly homogeneous under the condition of
being random.

\begin{acknowledgments}
The authors would like to express their gratitude to Markus
H\"utter for providing the colloidal microstructure data from the
Brownian dynamics simulations. Additionally, Iwan Schenker would
like to thank Henning Galinski and Joakim Reuteler for helpful
discussions.
\end{acknowledgments}

\newpage
\bibliography{Schenker_PhysRevE_2009}

\end{document}